\documentclass[12pt]{article}
\usepackage{graphicx}
 
\def \bea{\begin{eqnarray}}
\def \beq{\begin{equation}}
\def \eea{\end{eqnarray}}
\def \eeq{\end{equation}}
 
\def \ok{\overline{K}^0} 
\def \s{\sqrt{2}} 
\def \st{\sqrt{3}}

\begin{document} 
\rightline{EFI 18-2} 
\rightline{TECHNION-PH-2018-3}
\rightline{arXiv:1803.02267} 
\rightline{March 2018} 
\bigskip
\centerline{\bf AN OVERVIEW OF $\Lambda_c$ DECAYS} 
\bigskip
 
\centerline{Michael Gronau\footnote{gronau@physics.technion.ac.il}} 
\centerline{\it Physics Department, Technion -- Israel Institute of Technology} 
\centerline{\it 32000 Haifa, Israel} 
\medskip 
 
\centerline{Jonathan L. Rosner\footnote{rosner@hep.uchicago.edu}} 
\centerline{\it Enrico Fermi Institute and Department of Physics} 
\centerline{\it University of Chicago, 5640 S. Ellis Avenue, Chicago, IL 60637} 
 
\begin{quote}
The decays of the ground-state charmed baryon $\Lambda_c$ are now close to
being completely mapped out.  In this paper we discuss some remaining open
questions, whose answers can help shed light on weak processes contributing
to those decays, on calculations of such quantities as transition form factors
in lattice QCD, and on missing decay modes such as $\Lambda_c \to \Lambda^*
\ell^+ \nu_\ell$, where $\Lambda^*$ is an excited resonance.  The discussion is
in part a counterpart to a previous analysis of inclusive $D_s$ decays.
\end{quote} 

\leftline{PACS numbers: 14.20.Lq, 13.30.Eg, 12.40.Ee, 11.30.Ly} 
\bigskip 
 
\section{INTRODUCTION \label{sec:int}} 

The lowest-lying charmed baryon $\Lambda_c$ was discovered more than 40
years ago \cite{Knapp:1976et} , but its decays have not yet been fully
mapped out due to the many available modes.  Significant progress toward
this goal has been made in the past few years, thanks to advances in
particle identification, tracking, and collider luminosity.  In the present
paper we identify some missing modes of interest, the questions associated
with them, and ways of filling the gaps in our knowledge. Some modes involving
neutrons cannot be identified directly, so one must resort to models such
as isospin statistical models \cite{statmods,Peshkin:1976kw,Gronau:2009vt}.
These techniques also apply to modes with many neutral pions.  Even when
isospin multiplets have been filled, however, there remains a gap.  Some of
this gap arises from unreported modes with $\eta$ or $\eta'$.  In addition, we
propose that some of it be filled with semileptonic decays $\Lambda_c \to
\Lambda^* \ell^+ \nu_\ell$, where $\Lambda^*$ is either an excited resonance
such as $\Lambda(1405)$ or $\Lambda(1520)$ \cite{CWUpd}
number in parentheses denotes the mass in MeV) or a continuum $I=0$ state such
as $\Sigma \pi$ or $N \overline{K}$.

A global analysis of $\Lambda_c$ decays is particulary timely now that Belle
\cite{Zupanc:2013iki} and BESIII \cite{Ablikim:2015flg} have significantly
improved the accuracy of the branching fraction ${\cal B}(\Lambda_c \to p K^-
\pi^+)$, which has been used to normalize other $\Lambda_c$ branching fractions.
Their results and the resulting Particle Data group's \cite{PDG} ``fit'' value
are summarized in Table \ref{tab:pkpi}. BESIII \cite{Ablikim:2015flg} quotes
updated absolute branching fractions for a dozen $\Lambda_c$ modes,
incorporated into the latest PDG averages \cite{CWUpd}.  Also new are 
a set of updated branching fractions for $\Lambda_c \to \Sigma \pi \pi$,
including the first observation of the mode $\Lambda_c \to \Sigma^+\pi^0 \pi^0$
\cite{Berger:2018pli}.  The situation has greatly improved in the
past six years since a plea was issued for improvement of $\Lambda_c$ absolute
branching fractions \cite{Rosner:2012gj}.  The present paper is devoted in
part to an update of that analysis.  For an early model-dependent discussion of 
Cabibbo-favored two-body $\Lambda_c$ decays and for a recent study of 
singly Cabibbo-suppressed decays, quoting other papers using a similar 
approach, see Refs.\ \cite{Cheng:1993gf} and \cite{Cheng:2018hwl}, respectively.

\begin{table}
\caption{Values of ${\cal B}(\Lambda_c \to p K^- \pi^+)$.
\label{tab:pkpi}}
\begin{center}
\begin{tabular}{c c c} \\ \hline \hline
Source & Ref.\ & Value (\%) \\ \hline
Belle & \cite{Zupanc:2013iki} & $6.84\pm0.24^{+0.21}_{-0.27}$ \\
BESIII & \cite{Ablikim:2015flg} & $5.84\pm0.27\pm 0.23$ \\
PDG ``fit'' & \cite{CWUpd} & $6.23 \pm 0.33$ \\ \hline \hline
\end{tabular}
\end{center}
\end{table}

We review the isospin statistical method in Sec.\ \ref{sec:stat}, giving
examples of its predictions for the $N \overline{K} \pi$ modes in Sec.\
\ref{sec:NKP}, for the $\Sigma 2\pi$ modes in Sec.\ \ref{sec:S2P}, for
the $N \overline{K} 2\pi$ modes in Sec.\ \ref{sec:NK2P}, and for the
$\Sigma 3\pi$ modes in Sec.\ \ref{sec:S3P}.  Some other modes are treated in
Sec.\ \ref{sec:other}.  We apply the method to identify
missing charge modes in $\Lambda_c$ final states in
Sec.\ \ref{sec:miss}.  This approach mirrors one applied to $D_s$ decays
\cite{Gronau:2009vt}.  Section \ref{sec:imp} is devoted to suggestions for
placing these estimates on a firmer footing, such as identifying missing
neutrons and taking account of decays involving $\eta$ and $\eta'$.
Section \ref{sec:iso} is devoted to systematic errors associated
with possible deviations from the statistical isospin model, such as the
dominance of resonant substructure.  Part of
the remaining shortfall is proposed in Sec.\ \ref{sec:sld} to be filled
by $\Lambda_c$ semileptonic decays to excited states.  Section \ref{sec:con}
concludes.  An Appendix discusses details of obtaining branching fractions
not quoted by Ref.\ \cite{PDG}.

\section{STATISTICAL ISOSPIN MODEL \label{sec:stat}}

A multiparticle amplitude may be decomposed into a series of invariant isospin
amplitudes depending on particle momenta, 
as we shall show by examples in the next three sections.
The statistical isospin model \cite{statmods,Peshkin:1976kw} parametrizes
one's ignorance of underlying dynamics by assuming that each invariant
amplitude contributes equally and incoherently to each decay mode, with
relative branching fractions determined only by Clebsch-Gordan coefficients.
The squares of each invariant amplitude's coefficients then sum to 1,
and for each mode the branching fraction is the sum of squares of each
contributing amplitude, divided by the number of invariant amplitudes.
The answer does not depend on how the isospin decomposition is performed.

We illustrate this process for a three-body final state $ABC$ produced in a
state of definite isospin $I$ and third component $I_3$.  We may first
decompose the $BC$ system into isospin amplitudes $I^{BC}$ with $I^B - I^C
\le I^{BC} \le I^B + I^C$.  We then combine the amplitudes $I^{BC}$ with
$I^A$ in such a way that the final isospin is the desired value $I$, while
$I_3^{BC} + I_3^A = I_3$.  This process then reduces to manipulation of
Clebsch-Gordan coefficients.

\section{DECAYS $\Lambda_c \to N \overline{K} \pi$ \label{sec:NKP}}

The normalizing branching fraction for many $\Lambda_c$ decays is ${\cal B}
(\Lambda_c \to p K^- \pi^+)$.  The isospin-partner modes are $n \ok \pi^+$
and $p \ok \pi^0$.  The initial $\Lambda_c$ has isospin zero, while the
$\Delta S = 1$ (Cabibbo-favored) transition is governed by $c \to s u \bar
d$, resuting in a final state with $I = I_3 = 1$.  The final $N \overline{K}$
final state can have isospin zero or one.  If invariant amplitudes $A$ are
labeled by this isospin, one finds
\beq \label{eqn:nkpamps}
{\cal A}(p K^- \pi^+) = \frac{A_0}{\s} - \frac{A_1}{2}~,~~
{\cal A}(n \ok \pi^+) = - \frac{A_0}{\s} - \frac{A_1}{2}~,~~
{\cal A}(p \ok \pi^0) = \frac{A_1}{\s}~,
\eeq
satisfying the sum rule
\beq
{\cal A}(p K^- \pi^+) + {\cal A}(n \ok \pi^+) + \s\,{\cal A}(p \ok \pi^0) = 0
\eeq
as noted by BESIII, the observers of the $n \ok \pi^+$ mode
\cite{Ablikim:2016mcr}.  The statistical isospin model postulates equality
and incoherence of $A_0$ and $A_1$, so that the branching fractions of
$\Lambda_c$ to the above $N \overline{K} \pi$ modes are in the ratio
3/8:3/8:1/4.  This prediction is compared with data \cite{PDG} in Table
\ref{tab:NKP}.  The ratios of branching fractions for the two modes
with a proton are underestimated by $2.4\sigma$ and $1.5\sigma$ with respect
to measurements, while the branching fraction for the $n \ok \pi^+$ mode is
slightly overestimated by $2.9\sigma$.  This gives an idea of the degree
to which we can trust the statistical model.  Deviations from its predictions
will be discussed in Sec.\ \ref{sec:iso}.

\begin{table}
\caption{Statistical isospin model predictions for relative branching
fractions of $\Lambda_c$ to $N \overline{K} \pi$ final states and
comparison with observation.
\label{tab:NKP}}
\begin{center}
\begin{tabular}{c c c c} \hline \hline
     Final    &  Observed $\Lambda_c$   &     Fraction of      & Statistical \\
     state    & branching fraction (\%) & $N \overline{K} \pi$ & model\\ \hline
$p K^- \pi^+$ &    $6.23 \pm 0.33$      &  $0.452 \pm 0.032$   & 0.375 \\
$n \ok \pi^+$ &    $3.64 \pm 0.50$      &  $0.264 \pm 0.038$   & 0.375 \\
$p \ok \pi^0$ &    $3.92 \pm 0.26$      &  $0.284 \pm 0.023$   & 0.250 \\
Total $N \overline{K} \pi$ & $13.79\pm0.65$ & & \\ \hline \hline
\end{tabular}
\end{center}
\end{table}

One could, if desired, decompose the final states into ones labeled by the
isospin of the $\overline{K} \pi$ system, with invariant amplitudes $A_{1/2}$
and $A_{3/2}$.  Assuming these two amplitudes are equal in magnitude and
incoherent, one arrives at the same result.

\section{DECAYS $\Lambda_c \to \Sigma \pi \pi$ \label{sec:S2P}}

The final state in $\Lambda_c \to \Sigma \pi \pi$ decays must have $I=I_3=1$,
as noted in the previous subsection.  One way to count invariant amplitudes is
to designate them by the isospin of the two-pion system, $I_{\pi \pi} = 0, 1,
2$.  For each such isospin there is a unique coupling with the $\Sigma$
(whose isospin is 1) to the $I=1$ final state.  Thus there are three
invariant amplitudes $A_0$, $A_1$, $A_2$.  The $\Lambda_c$ decay amplitudes are
expressed in terms of them as:
\bea
{\cal A}(\Sigma^- \pi^+ \pi^+) & = & \sqrt{\frac35} A_2~,\\
{\cal A}(\Sigma^0 \pi^+ \pi^0) & = & -\frac12 \sqrt{\frac35} A_2
 + \frac12 A_1~, \\
{\cal A}(\Sigma^0 \pi^0 \pi^+) & = & -\frac12 \sqrt{\frac35} A_2
 - \frac12 A_1~, \\
{\cal A}(\Sigma^+ \pi^+ \pi^-) & = & \frac{1}{2\sqrt{15}} A_2 - \frac12 A_1
 + \frac{1}{\st} A_0~, \\
{\cal A}(\Sigma^+ \pi^- \pi^+) & = & \frac{1}{2\sqrt{15}} A_2 + \frac12 A_1
 + \frac{1}{\st} A_0~, \\
{\cal A}(\Sigma^+ \pi^0 \pi^0) & = &  \frac{1}{\sqrt{15}} A_2 
 - \frac{1}{\st} A_0~.
\eea
Here we quote amplitudes for both orders of differing pion charges, needed
for the sum of squares of coefficients of each isospin amplitude to add up
to 1.  The statistical-model predictions are obtained by assuming that
invariant amplitudes are equal in magnitude and incoherent. They are compared
with experiment \cite{PDG,Berger:2018pli} in Table \ref{tab:S2P}.  The
parentheses around pion pairs denote the sum of both orders.  The agreement
with the statistical isospin model is quite good.

\begin{table}
\caption{Statistical isospin model predictions for relative branching
fractions of $\Lambda_c$ to $\Sigma \pi  \pi$ final states and
comparison with observation.  
\label{tab:S2P}}
\begin{center}
\begin{tabular}{c c c c} \hline \hline
     Final    &  Observed $\Lambda_c$   &     Fraction of      & Statistical \\
     state    & branching fraction (\%) & $\Sigma \pi \pi$ & model\\ \hline
$\Sigma^- \pi^+ \pi^+$ & $1.86 \pm 0.18$ & $0.177 \pm 0.018$ & 0.200 \\
$\Sigma^0(\pi^+\pi^0)$ & $3.03 \pm 0.23$ & $0.288 \pm 0.024$ & 0.267 \\
$\Sigma^+(\pi^+\pi^-)$ & $4.41 \pm 0.20$ & $0.419 \pm 0.024$ & 0.400 \\
$\Sigma^+ \pi^0 \pi^0$ & $1.23 \pm 0.12$ & $0.117 \pm 0.012$ & 0.133 \\
Total $\Sigma \pi \pi$ & $10.53 \pm 0.37$ & & \\ \hline \hline
\end{tabular}
\end{center}
\end{table}

In Ref.\ \cite{Peshkin:1976kw} bounds were placed on $\Lambda_c$ decays
to $\Sigma \pi \pi$ final states consisting of all charged particles, with
the result
\beq \label{eqn:spp}
\frac{1}{2} \le \frac{{\cal B}(\Lambda_c \to \Sigma^- \pi^+ \pi^+)
  + {\cal B}(\Lambda_c \to \Sigma^+ (\pi^+ \pi^-))}
{{\cal B}(\Lambda_c \to \Sigma \pi \pi)} \le \frac{4}{5}~.
\eeq
The quotient in Eq.\ (\ref{eqn:spp}) has the value 3/5 in the statistical
isospin model.

\section{DECAYS $\Lambda_c \to N \overline{K} \pi \pi$ \label{sec:NK2P}}

One convenient way to define invariant amplitudes for the $N \overline{K}
\pi \pi$ final state is to couple the $N \overline{K}$ pair to isospin
$I_{N\overline{K}} =0,1$ and the pion pair to $I_{\pi\pi}=0,1,2$.  When 
$I_{N\overline{K}}=0$, only $I_{\pi\pi} = 1$ can lead to a final state with
$I=1$; we call the corresponding reduced amplitude $A_{1a}$.  When
$I_{N\overline{K}}= 1$, the $I=1$
final state receives contributions from $I_{\pi\pi}=0,1,2$; we call the
corresponding reduced amplitudes $A_0, A_{1b}, A_2$.  The decomposition of
$\Lambda_c$ decay amplitudes in terms of these reduced amplitudes is
\bea
{\cal A}(p K^- \pi^+ \pi^0) & = & -\frac12\sqrt{\frac{3}{10}} A_2
 + \frac12 A_{1a} + \frac{1}{2\s} A_{1b}~,\\
{\cal A}(p K^- \pi^0 \pi^+) & = & -\frac12\sqrt{\frac{3}{10}} A_2 
 - \frac12 A_{1a} - \frac{1}{2\s} A_{1b}~,\\
{\cal A}(n \ok \pi^+ \pi^0) & = & -\frac12\sqrt{\frac{3}{10}} A_2 
 - \frac12 A_{1a} + \frac{1}{2\s} A_{1b}~,\\
{\cal A}(n \ok \pi^0 \pi^+) & = & -\frac12\sqrt{\frac{3}{10}} A_2 
 + \frac12 A_{1a} - \frac{1}{2\s} A_{1b}~,\\
{\cal A}(p \ok \pi^0 \pi^0) & = & \frac{1}{\sqrt{15}} A_2
 - \frac{1}{\st} A_0~,\\
{\cal A}(p \ok \pi^+ \pi^-) & = & \frac{1}{2\sqrt{15}} A_2
 - \frac12 A_{1b} + \frac{1}{\st} A_0~,\\
{\cal A}(p \ok \pi^- \pi^+) & = & \frac{1}{2\sqrt{15}} A_2
 + \frac12 A_{1b} + \frac{1}{\st} A_0~,\\
{\cal A}(n K^- \pi^+ \pi^+) & = & \sqrt{\frac35} A_2~.
\eea
The implications of this decomposition for the statistical isospin model
are summarized in Table \ref{tab:NK2P}.  Only two modes are measured, and they
imply quite different values of the total ${\cal B}(\Lambda_c \to N
\overline{K} \pi \pi)$.  We shall bear this uncertainty in mind when evaluating
the accuracy of our predictions.  Using the average value of the total
branching fraction, we predict the branching fractions for as yet unseen
decay modes in brackets in the table.

\begin{table}
\caption{Statistical isospin model predictions for relative branching
fractions of $\Lambda_c$ to $N \overline{K} \pi \pi$ final states and
comparison with observation.  Quantities in brackets are inferred from
implied total average.
\label{tab:NK2P}}
\begin{center}
\begin{tabular}{c c c c} \hline \hline
        Final        &  Statistical & $\Lambda_c$ branching & Implied total \\
        state        &     model    &  fraction (\%)  &
 ${\cal B}(\Lambda_c \to N \overline{K} \pi \pi)$ (\%) \\ \hline
$p K^- (\pi^+\pi^0)$ & 9/40 = 0.225 & $4.42 \pm 0.31$ & $19.64 \pm 1.38$ \\
$n \ok (\pi^+\pi^0)$ & 9/40 = 0.225 & [$3.07\pm0.16$] & -- \\
 $p \ok \pi^0 \pi^0$ & 1/10 = 0.100 & [$1.36\pm0.07$] & -- \\
$p \ok (\pi^+\pi^-)$ & 3/10 = 0.300 & $3.18 \pm 0.24$ & $10.60 \pm 0.80$ \\
 $n K^- \pi^+ \pi^+$ & 3/20 = 0.150 & [$2.05\pm0.11$] & -- \\
Average              &              &                 & $12.88 \pm 0.69$ (a) \\
\hline \hline
\end{tabular}
\end{center}
\leftline{(a) Error to be multiplied by a scale factor of 5.67 in the final
total.}
\end{table}

The average of implied total branching fractions involves two very disparate
values, so we apply a scale factor of 5.67 to the error, giving 3.92 to
be quoted in the summary table.  Possible deviations from the statistical
isospin model will be noted in Sec.\ \ref{sec:iso}.

\section{DECAYS $\Lambda_c \to \Sigma 3\pi$ \label{sec:S3P}}

The number of invariant amplitudes may be counted by noting the number of
$3 \pi$ amplitudes with each isospin and then coupling them up with the $I = 1$ $\Sigma$ to a final state with $I=1$.  The multiplicities of three-pion
amplitudes are 1 for $I=0$, 3 for $I = 1$, 2 for $I = 2$, and 1 for $I = 3$.
Each of these except the $I = 3$ amplitude can couple up with the $\Sigma$
to form final isospin 1.  Thus there are a total of six reduced amplitudes.

The statistical model's predictions of relative branching fractions for
Cabibbo-favored decays have been given in Ref.\ \cite{Peshkin:1976kw}.  For
$\Sigma 3 \pi$ final states of $\Lambda_c$ we show the results in Table
\ref{tab:S3P}.  In averaging the two values leading to different implied total
fractions we multiply the uncertainty of 1.2\% by a scale factor of 2 to give a
final uncertainty of 2.4\%.  Deviations from the statistical isospin model will
be discussed in Sec.\ \ref{sec:iso}.  Bounds on $\Lambda_c$ decays to $\Sigma
3\pi$ final states with three charged particles \cite{Peshkin:1976kw} are

\beq \label{eqn:s3p}
\frac{3}{5} \le \frac{{\cal B}(\Lambda_c \to \Sigma^- 2\pi^+ \pi^0)
  + {\cal B}(\Lambda_c \to \Sigma^+ \pi^+\pi^-\pi^0)
  + {\cal B}(\Lambda_c \to \Sigma^0 2\pi^+ \pi^-)}
{{\cal B}(\Lambda_c \to \Sigma 3\pi)} \le 1~.
\eeq
The quotient in Eq.\ (\ref{eqn:s3p}) has the value 4/5 in the statistical
isospin model.

\begin{table}
\caption{Statistical model predictions and observed branching fractions for
$\Lambda_c \to \Sigma 3 \pi$ decays.  For each mode, all permutations of pions
are implied.  
\label{tab:S3P}}
\begin{center}
\begin{tabular}{c c c c} \hline \hline
        Final        &  Statistical & $\Lambda_c$ branching & Implied total \\
        state        &     model    &  fraction (\%)  &
 ${\cal B}(\Lambda_c \to \Sigma 3 \pi$) (\%) \\ \hline
$\Sigma^0 2\pi^+\pi^-$ & 1/5 = 0.200 & $1.10 \pm 0.30$ & $5.5 \pm 1.5$ \\
$\Sigma^-\pi^0 2\pi^+$ & 1/5 = 0.200 & $2.1 \pm 0.4$ & $10.5 \pm 2.0$ \\
$\Sigma^+\pi^+\pi^-\pi^0$ & 2/5 = 0.400 & (a) & -- \\
$\Sigma^0 \pi^+ 2\pi^0$ & 3/20 = 0.150 & -- & -- \\
   $\Sigma^+ 3\pi^0$ & 1/20 = 0.050 & -- & -- \\
        Average      &              &    & $7.3 \pm 1.2$ (b) \\ \hline \hline
\end{tabular}
\end{center}
\leftline{(a) $B(\Lambda_c \to \Sigma^+ \omega) = (1.69 \pm 0.21)\%$ counted
separately}
\leftline{(b) Error to be multiplied by a scale factor of 2.0 in the final
total.}
\end{table}

\section{SOME OTHER CABIBBO-FAVORED MODES \label{sec:other}}

We use a standard format, extracting estimates of branching fraction to
the sum of all charge states for a given mode and averaging where there is
more than one measured charge state.  The statistical model fractions are
taken from Table 4 of Ref.\ \cite{Peshkin:1976kw}.

\subsection{$N \overline{K} 3\pi$}

Only one mode ($K^-p 2\pi^+ \pi^-$) is used in estimating the total, as the
branching fraction for $K^- p \pi^+ 2\pi^0$ is suspiciously large in
comparison with the all-charged-particle mode.  It bears watching, however.
(See Table \ref{tab:NK3P}.)

\subsection{$\Lambda 3\pi$}

So far only the mode with no neutral pions has been detected.
Ref.\ \cite{Peshkin:1976kw} obtains the bounds
\beq
\frac{1}{2} \le \frac{{\cal B}(\Lambda_c \to \Lambda 2 \pi^+ \pi^-)}
{{\cal B}(\Lambda_c \to 3 \pi)} \le \frac{4}{5}~,
\eeq
where the value of the quotient in the statistical isospin model is 3/5
(see Table \ref{tab:L3P}).

\subsection{$\Lambda 4 \pi$}

The mode with a single neutral pion is the only one detected.  With three
neutral pions the missing mode is unlikely to be confirmed soon.
Ref.\ \cite{Peshkin:1976kw} finds the bounds
\beq
\frac{3}{5} \le \frac{{\cal B}(\Lambda_c) \to \Lambda \pi^- \pi^0 2\pi^+)}
{{\cal B}(\Lambda_c \to \Lambda 4 \pi)} \le 1~,
\eeq
with the statistical isospin model giving 4/5 for the quotient (see Table
\ref{tab:L4P}).

\begin{table}
\caption{Statistical model predictions and observed branching fractions for
$\Lambda_c \to N \overline{K} 3 \pi$ decays.  For each mode, all permutations
of pions are implied.
\label{tab:NK3P}}
\begin{center}
\begin{tabular}{c c c c}
        Final        &  Statistical & $\Lambda_c$ branching & Implied total \\
        state        &     model    &  fraction (\%)  &
 ${\cal B}(\Lambda_c \to N \overline{K} 3\pi) (\%)$ \\ \hline
$K^-p 2\pi^+ \pi^-$  &  1/6 = 0.167  & $0.14\pm0.09$ & $0.84 \pm 0.54$ \\
$\ok n 2\pi^+ \pi^-$ &  1/6 = 0.167  & -- & -- \\
$\ok p\pi^+\pi^-\pi^0$ & 4/15 = 0.267 & -- & -- \\
$K^- n 2\pi^+ \pi^0$ & 2/15 = 0.133 & -- & -- \\
$K^- p \pi^+ 2\pi^0$ & 7/60 = 0.117 & (a) & -- \\
$\ok n \pi^+ 2\pi^0$ & 7/60 = 0.117 & -- & -- \\
    $\ok p 3\pi^0$   & 1/30 = 0.033 & -- & -- \\
   Average           &              &    & $0.84 \pm 0.54$ \\ \hline \hline
\end{tabular}
\end{center}
\leftline{(a) The PDG value of $1.0\pm0.5$ is ignored but bears watching.}
\end{table}

\begin{table}
\caption{Statistical model predictions and observed branching fractions for
$\Lambda_c \to \Lambda 3 \pi$ decays. For each mode, all permutations
of pions are implied.
\label{tab:L3P}}
\begin{center}
\begin{tabular}{c c c c}
        Final        &  Statistical & $\Lambda_c$ branching & Implied total \\
        state        &     model    &  fraction (\%)  &
 ${\cal B}(\Lambda_c \to \Lambda 3\pi)(\%)$ \\ \hline
$\Lambda \pi^- 2\pi^+$ & 3/5 = 0.600 & $3.61\pm0.29$ & $6.02 \pm 0.48$ \\
$\Lambda \pi^+ 2\pi^0$ & 2/5 = 0.400 & -- & -- \\
    Average          &               &    & $6.02 \pm 0.48$ \\ \hline \hline
\end{tabular}
\end{center}
\end{table}

\begin{table}
\caption{Statistical model predictions and observed branching fractions for
$\Lambda_c \to \Lambda 4\pi$ decays.  For each mode, all permutations
of pions are implied.
\label{tab:L4P}}
\begin{center}
\begin{tabular}{c c c c}
        Final        &  Statistical & $\Lambda_c$ branching & Implied total \\
        state        &     model    &  fraction (\%)  &
 ${\cal B}(\Lambda_c \to \Lambda 4\pi) (\%)$ \\ \hline
$\Lambda\pi^-\pi^02\pi^+$ & 4/5 = 0.800 & $2.2 \pm 0.8$ & 
$2.75 \pm 1.00$ \\
 $\Lambda \pi^+ 3\pi^0$ & 1/5 = 0.200 & -- & -- \\
    Average         &               &      & 
 $2.75 \pm 1.00$ \\ \hline \hline 
\end{tabular}
\end{center}
\end{table}

\section{IDENTIFYING MISSING MODES \label{sec:miss}}

In the previous sections we have used the isospin statistical model to
estimate missing charge states for $\Lambda_c$ decay modes due to the
Cabibbo-favored process $c \to s u \bar d$, populating final states
with strangeness $S=-1$ and isospin $I = I_3 = 1$.  The results are
shown in Table \ref{tab:brs}.  Also shown are much rougher estimates
of branching fractions to $S=0$ final states, populated by the
singly-Cabibbo-suppressed transitions $c\to d u \bar d$ and $c\to s u \bar s$.

The sum of the two sets of branching fractions is $(89.6 \pm 5.0)\%$.  Thus
there is a hint, though not statistically compelling at present, that about
10\% of $\Lambda_c$ decays remain to be accounted for.  We shall suggest
that this could be due to semileptonic $\Lambda_c$ decays to excited
states such as $\Lambda(1405)$, $\Lambda(1520)$, or continuum $\Sigma \pi$
and/or $N \overline{K}$ states.

The estimates for the $\Delta S = 0$ transitions are very rough, as many
of them rely on the assumption that each charge mode is equally populated.
What one sees in the statistical model, instead, is that the modes with
the most neutral pions tend to be populated the least.  Thus the total
branching fraction for $\Delta S = 0$ decays may in fact be an upper bound.

A recent BESIII determination of inclusive $\Lambda$ production in $\Lambda_c$
decays \cite{Ablikim:2018jfs} finds ${\cal B}(\Lambda_c \to \Lambda + X) =
(38.2^{+2.8}_{-2.2} \pm 0.8)\%$.  We can compare this result with the sum
of contributing entries in Table \ref{tab:brs}.  Table \ref{tab:lams} shows
the final states directly leading to a $\Lambda$, and separately gives those
leading to a $\Sigma^0$, which decays 100\% of the time to $\Lambda \gamma$.
The sum of these totals is $(31.72 \pm 1.44)\%$, a shortfall of $2.4 \sigma$.
What could fill the gap?  Possible candidates are underestimates of modes
$\Lambda n \pi~(n=3,4)$ using the statistical model, modes $\Lambda n \pi~(n >
4)$ or $\Sigma^0 n \pi~(n > 3)$, and semileptonic decays to hadronic final
states consisting of $\Lambda$ accompanied by other particles.  Examples are
$\Lambda(1405,1520) \to \Sigma^0 \pi^0 \to \Lambda \gamma \pi^0$,
$\Lambda(1690) \to \Lambda 2 \pi$, and $\Lambda$ in nonresonant continuum.

\begin{table}
\caption{Observed and extrapolated branching fractions ${\cal B}$ for
$\Lambda_c$ decays, in \%.  Unless shown otherwise, the statistical isospin
model has been used to extrapolate to unseen charge states.
\label{tab:brs}}
\begin{center}
\begin{tabular}{c c c c} \hline \hline
\multicolumn{2}{c}{$\Delta S = -1$ transitions} &
\multicolumn{2}{c}{$\Delta S = 0$ transitions} \\
Mode & ${\cal B}$ & Mode & ${\cal B}$ \\
       $p \ok$     & $3.16 \pm 0.16$ & $p \eta$ & $0.124 \pm 0.030$  \\
$N \overline{K}\pi$ & $13.79\pm0.65$ & $N\pi\pi$ & $1.26\pm0.12$ (a) \\
   $p \ok \eta$    &  $1.6 \pm 0.4$  &  $N3\pi$  & $1.22 \pm 0.30$ (b) \\
$N\overline{K}2\pi$ & $12.88\pm 3.92$ &  $N4\pi$  & $1.10 \pm 0.70$ (a) \\
$N\overline{K}3\pi$ & $0.84\pm 0.54$ & $NK\overline{K}$ & $0.30\pm0.12$ (a)\\
  $\Lambda \pi^+$  & $1.29 \pm 0.07$ & $\Lambda K^+$ & $0.06 \pm 0.012$ \\
$\Lambda\pi^+\pi^0$ &  $7.0 \pm 0.4$ &   $\Sigma K$  & $0.102\pm0.016$ (a) \\
 $\Lambda 3\pi$  & $6.02 \pm 0.48$ & $\Sigma K \pi$ & $1.05\pm0.30$ (a) \\
 $\Lambda 4\pi$  & $2.75 \pm 1.00$ & $n e^+ \nu_e$ & $0.41 \pm 0.03$ (c) \\
 $\Sigma \pi$   & $2.52 \pm 0.12$ & $n\mu^+\nu_\mu$ & $0.40\pm0.03$ (c) \\
 $\Sigma \eta$   & $0.69 \pm 0.23$ & $p \pi^0$ & 0.008 (d) \\
 $\Sigma 2\pi$   & $10.53\pm0.37$  & $n \pi^+$ & 0.027 (d) \\
 $\Sigma 3\pi$   & $7.3 \pm 2.4$   & & \\
 $\Sigma \omega$  & $1.69 \pm 0.21$ & & \\
$\Lambda K^+ \ok$ & $0.56 \pm 0.11$ & & \\
$\Sigma K \overline{K}$ & $1.36\pm0.16$ (a) & & \\
 $\Xi^0 K^+$   & $0.55 \pm 0.07$ (e) & & \\
 $\Xi K \pi$    & $1.86\pm0.18$ (f) & & \\
 $\Lambda e^+\nu_e$ & $3.63 \pm 0.43$ (g) & & \\
$\Lambda\mu^+\nu_\mu$ & $3.49\pm0.53$ (h) & & \\
Total $\Delta S=-1$ & $83.51\pm4.92$ & Total $\Delta S = 0$ & $6.06\pm0.84$  \\
 \hline\hline
\end{tabular}
\end{center}
\leftline{(a) Branching fraction for one observed charge mode multiplied
 by number of charge states.}
\leftline{(b) Branching fraction to $p \pi^+ \pi^0 \pi^-$ taken as $(0.304
\pm 0.076)\%$ (geometric mean of $p \pi^+ \pi^-$}
\leftline{~~and $p 2\pi^+ 2 \pi^-$ modes), multiplied by 4 for total number of
charge states.}
\leftline{(c) Lattice QCD calculation \cite{Meinel:2017ggx}.
(d) Theoretical estimate from Ref.\ \cite{Cheng:2018hwl}.}
\leftline{(e) New value of $(0.59 \pm 0.09)\%$ \cite{Ablikim:2018bir}
  averaged with PDG value $(0.49 \pm 0.12)\%$ \cite{PDG}.}
\leftline{(f) We multiply ${\cal B}(\Lambda_c \to \Xi^- K^+ \pi^+)=
(0.62 \pm 0.06)\%$ \cite{PDG} by three to include charge states}
\leftline{~~$\Xi^0 K^+ \pi^0$ and $\Xi^0 K^0 \pi^+$.
  Ref.\ \cite{Ablikim:2018bir} measures ${\cal B}(\Lambda_c \to
\Xi^0(1530) K^+) = (0.50 \pm 0.10)\%$,}
\leftline{~~accounting for part but not all of the $\Xi^- K^+ \pi^+$ final
state.
(g) Ref.\ \cite{Ablikim:2015prg} (h) Ref.\ \cite{Ablikim:2016vqd}.}
\end{table}

\begin{table}
\caption{Final states in $\Lambda_c$ decay leading directly to a $\Lambda$
(left column) or through a $\Sigma^0$ (right column)
\label{tab:lams}}
\begin{center}
\begin{tabular}{c c c c} \hline \hline
State & ${\cal B}(\%)$ & State & ${\cal B}(\%)$ \\ \hline
$\Lambda \pi^+$ & $1.29 \pm 0.07$ & $\Sigma^0 \pi^+$ & $1.28 \pm 0.07$ \\
$\Lambda \pi^+ \pi^0$ & $7.0\pm0.4$ & $\Sigma^0 \pi^+\pi^0$ & $3.03\pm0.23$ \\
$\Lambda 3\pi$ & $6.02\pm0.48$ & $\Sigma^0 \pi^- 2\pi^+$ & $1.10 \pm 0.30$ \\
$\Lambda 4\pi$ & $2.75\pm1.00$ & $\Sigma^0 \pi^+ 2\pi^0$ & $1.10\pm0.18$ (a)\\
$\Lambda K^+ \ok$ & $0.56\pm0.11$ & $\Sigma^0 K^+$ & $0.051 \pm 0.008$ \\
$\Lambda e^+\nu_e$ & $3.63\pm0.43$ & $\Sigma^0K^+\pi^0$ & $0.21\pm0.06$ (b) \\
$\Lambda\mu^+\nu_\mu$&$3.49\pm0.53$ & $\Sigma^0K^0\pi^+$ & $0.21\pm0.06$ (b) \\
Total & $24.74 \pm 1.37$ & Total & $6.98 \pm 0.43$ \\ \hline \hline
\end{tabular}
\end{center}
\leftline{(a) See Table \ref{tab:S3P}:  $(3/20)\cdot(7.3 \pm 1.2)$}
\leftline{(b) Assuming equal to measured ${\cal B}(\Lambda_c \to \Sigma^+ K^+
\pi^-)$ \cite{PDG}}
\end{table}

\section{SUGGESTIONS FOR IMPROVEMENT \label{sec:imp}}

\subsection{Modes needing further attention}

The uncertainty on the $\Delta S = -1$ transitions is dominated by the
disagreement with the isospin statistical model in the $N \overline{K}
\pi \pi$ modes.  The ratio ${\cal B}(\Lambda_c \to p K^- \pi^+ \pi^0)/
{\cal B}(\Lambda_c \to p \ok \pi^+ \pi^-)$ is measured to be $1.39 \pm 0.11$,
whereas in the isospin statistical model it is predicted to be (9/40)/(3/10)
= 3/4.  The corresponding total $N \overline{K} 2\pi$ branching ratio is
very different depending on which mode one uses to estimate it.  Measurement
of further $N \overline{K} 2\pi$ modes might help to resolve the ambiguity.

The singly-Cabibbo-suppressed ($\Delta S = 0$) modes are not readily
amenable to a statistical treatment, as the final states are a mixture of
$I = 1/2$ and $I =3/2$.  Thus measuring their branching fractions in the
widest possible cases is called for instead.

\subsection{Neutron identification}

BESIII has recently reported observation of the first $\Lambda_c$ mode
containing a neutron \cite{Ablikim:2016mcr}.  The method used was to
ensure production of a $\Lambda_c$ using a combination of single and
double tags at a center-of-mass energy in $e^+ e^-$ collisions just above
$\Lambda_c^+ \Lambda_c^-$ threshold.  The neutron was then inferred from
kinematic reconstruction.  In principle this method could be applied to
many states in the $N \overline{K} 2 \pi$ and $N \overline{K} 3 \pi$ modes.
The presence of a neutron in a kinematically constrained fit could be
confirmed if there were a calorimetric signal (resembling the
interaction of a $K^0_{\rm L}$) in the outer layer of a detector such as
BESIII or Belle.

\subsection{Inclusive $\eta,\eta'$ branching fractions}

Although some portion of decay modes involving $\eta$ or $\eta'$ appears
in multi-pion final states, the inclusive $\eta$ and $\eta'$ branching
fractions have not been reported.  It would be very helpful to have them,
in the same manner that inclusive measurements were very helpful in sorting
out $D_s$ decays \cite{Gronau:2009vt}.

\section{DEVIATIONS FROM STATISTICAL MODEL \label{sec:iso}}

In all decays involving three or more final-state particles, pairwise
associations in resonant substructures can lead to deviations from the
statistical isospin model.  However, in their high-statistics studies of
$\Lambda_c$ decays, neither BESIII \cite{Ablikim:2015flg} nor Belle
\cite{Berger:2018pli} show Dalitz plots or one-dimensional plots of pairwise
effective masses.  Consequently, we have to anticipate possible deviations from
the statistical isospin model without the help of experiment.  We hope this
situation will change in the near future.

In Figure 1 we give three examples of processes contributing
to $\Lambda_c$ decays.  These have the potential of populating final states
in a manner differing from the statistical isospin model, giving rise to
characteristic resonant substructures.  We shall estimate the corresponding
uncertainties for a series of final states.  In cases where all charge states
are allowed, we will comment on how well the statistical isospin model is
obeyed, but will not assign any uncertainty to the branching fractions.

\begin{figure}
\label{fig:procs}
\begin{center}
\includegraphics[width = 0.31\textwidth]{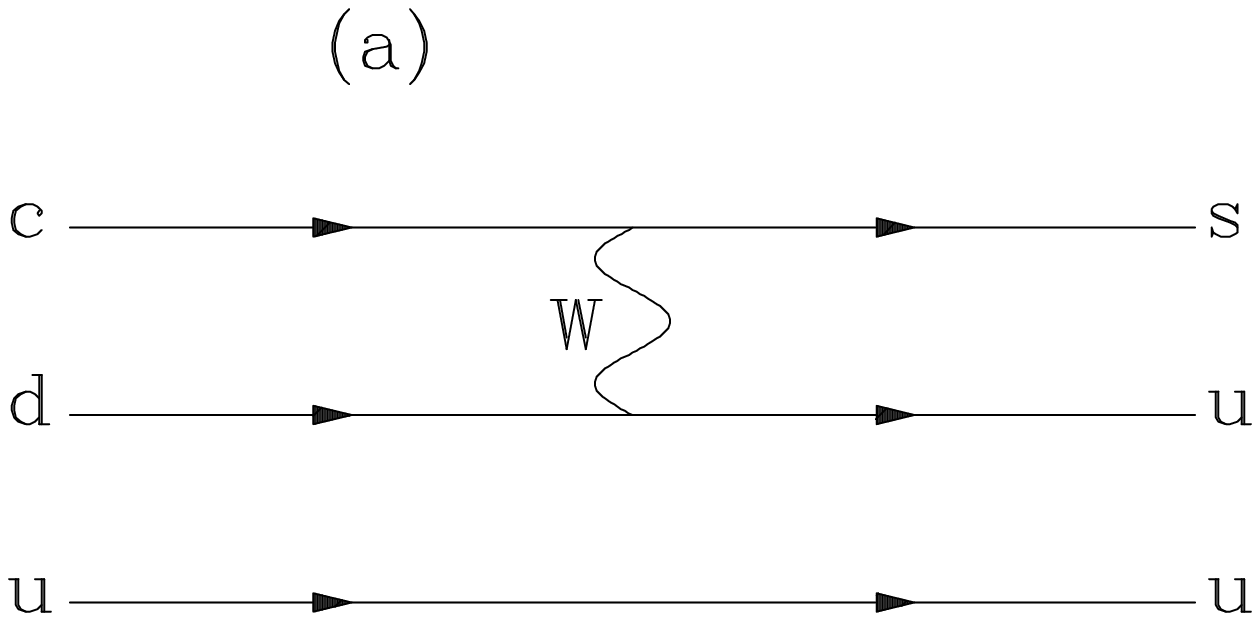} \hskip 0.1in
\includegraphics[width = 0.31\textwidth]{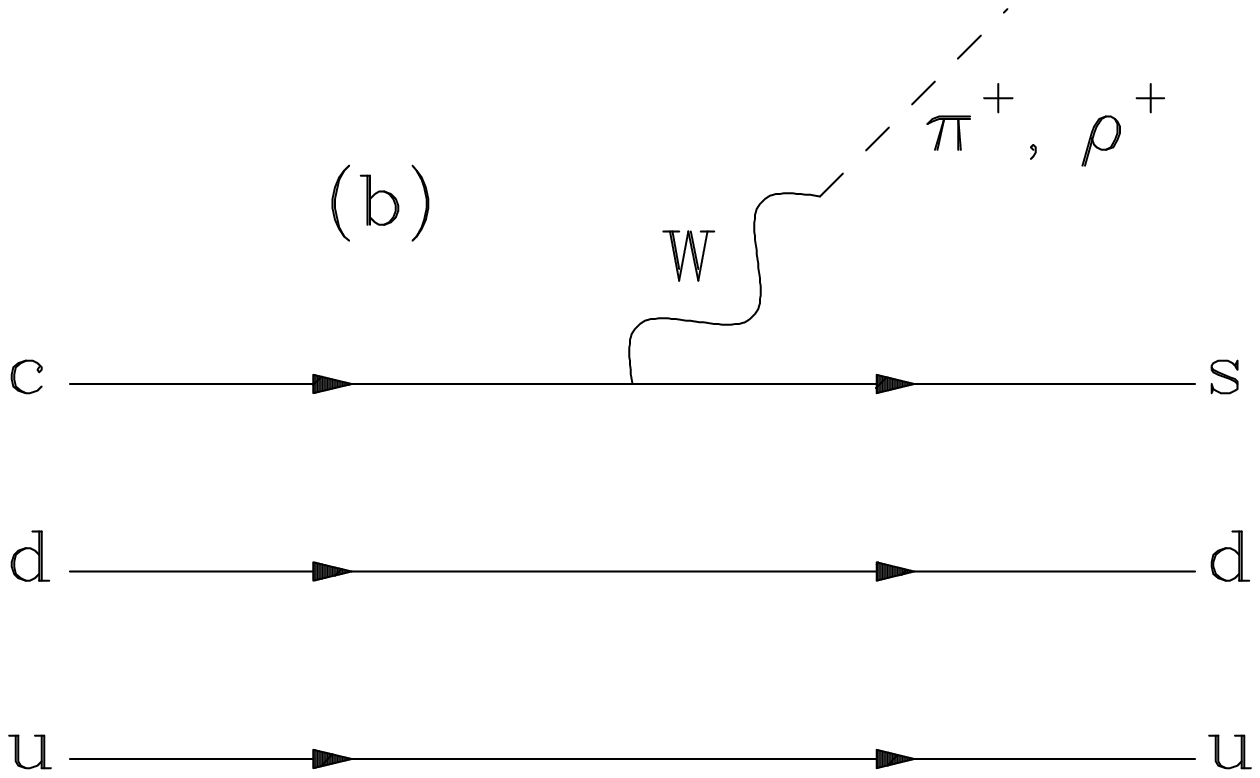} \hskip 0.1in
\includegraphics[width = 0.31\textwidth]{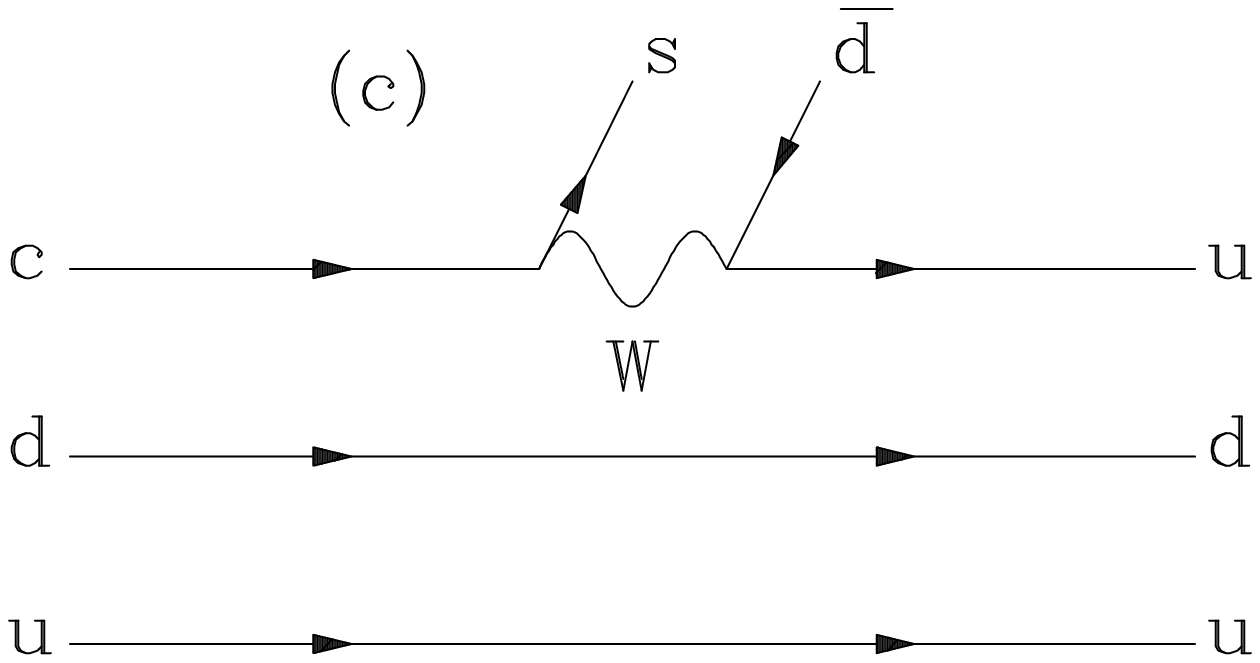}
\end{center}
\caption{Cabibbo-favored processes contributing to $\Lambda_c$ hadronic
decays.  (a) Internal conversion involving the subprocess $cd \to su$ with
$W$ exchange.  (b) Spectator process with $c \to (\pi^+,\rho^+) s$ with
isospin-zero $ud$ pair as a spectator.  (c) Color-suppressed process
involving $c \to (s \bar d) u$, where $s \bar d \to \bar K^0,\bar K^{*0},
\ldots$.  For Cabibbo-suppressed modes, replace $s$ with $d$.}
\end{figure}

\subsection{$N \overline{K} \pi$}

All branching fractions have been observed.  Defining $R_1 = {\cal B}
(p K^- \pi^+)$, $R_2 = {\cal B}(n \ok \pi^+)$, $R_3 = {\cal B} (p \ok \pi^0)$,
amplitudes in Eq.\ (\ref{eqn:nkpamps}) (up to a common factor) are related by
\bea
|A_0|^2 & = & R_1 + R_2 - 2 R_3 = 5.95 \pm 0.65~,\\
|A_1|^2 & = & 2 R_3 = 7.84 \pm 0.52~,\\
{\rm Re}(A_0^* A_1) & = & (R_2 - R_1)/\sqrt{2} = -1.83 \pm 0.42~.
\eea
The two amplitudes $A_0$ and $A_1$ are unequal in magnitude and have some
degree of coherence, in contrast to the statistical isospin model which
would have them equal in magnitude and out of phase with one another.

One can qualitatively anticipate the violation of the statistical isospin
model by reference to the three diagrams of Fig.\ 1.  The
relatively short lifetime of the $\Lambda_c$, about 0.2 ps \cite{PDG},
can be ascribed in large part to the contribution of Fig.\ 1(a),
which leads to a final state $suu$ resembling a $J=1/2$ excited $\Sigma^{*+}$.
This can hadronize by production of either a $u \bar u$ or $d \bar d$ pair.
In the former case one produces a configuration $(s \bar u)(uuu)$ which can
materialize as $K^- \Delta^{++} = K^-p\pi^+$.  The latter case leads to a
configuration $(s \bar d)(duu)$ which can materialize to $\ok n \pi^+$ or
$\ok p \pi^0$.  In this example, $K^- \pi^+$ accounts for 50\% of $\Lambda_c
\to N \overline{K} \pi$ decays, exhibiting the likely direction of deviation
from the statistical isospin model.

What if the spectator diagram (b) were dominant?  The spectator $ud$ pair would
remain in an isospin-zero final state, implying equal branching fractions
for $n \ok \pi^+$ and $p K^- \pi^+$, and no contribution to $p \ok \pi^0$,
far from the observed situation.

\subsection{$\Sigma \pi \pi$}

Here the statistical isospin model works surprisingly well.  Both the
internal conversion and spectator processes can contribute.  They both can
excite resonances such as $\Lambda(1405) (J^P = 1/2^-)$ and $\Lambda(1520)
(J^P = 3/2^-)$, which in turn can couple to all charge states of $\Sigma \pi$
and $N \overline{K}$.  (The contribution to the $N \overline{K}$ amplitude
of the $\Lambda(1405)$ would involve an intermediate off-shell state.)
Either hyperon resonance decays to an $I=0$ final state, so $\Lambda(1405)
\to (\pi^+ \Sigma^-),(\pi^0 \Sigma^0),(\pi^-\Sigma^+)$ in equal proportions.
The spectator process does not contribute to $\Lambda_c \to \Sigma^+\pi^0\pi^0$,
which is the smallest $\Sigma \pi \pi$ branching fraction, both predicted by
the isospin statistical model and observed.

\subsection{$N \overline{K} \pi \pi$}

The ratio of branching ratios to $p K^- \pi^+\pi^0$ and $p \ok \pi^+ \pi^-$
is $(4.42 \pm 0.31)/(3.18 \pm 0.24) = 1.39 \pm 0.14$, very far from the
statistical isospin model's prediction of $(9/40)/(3/10) = 3/4$.  The $p \ok
\pi^+ \pi^-$ amplitude would be suppressed if the $N \overline{K}$ amplitude
were predominantly $I=0$, as if dominated by $\Lambda(1405)$ and
$\Lambda(1520)$.  In that case the only nonzero amplitude in Sec.\
\ref{sec:NK2P} would be $A_{1a}$, and the mode $p K^- \pi^+ \pi^0$ would be
1/4 of the $N \overline{K} 2\pi$ total.  Thus for the $N \overline{K} 2\pi$
total one would have $4 \times (4.42 \pm 0.31)\% = (17.68 \pm 1.24)\%$. Taking
this value rather than $(12.88 \pm 0.69)\%$ on the bottom line of Table
\ref{tab:NK2P} one sees a difference of $4.80\%$, which could be substituted
for the scaled error of (0.69)(5.67) = 3.92\% ascribed to this mode.

\subsection{$\Sigma 3\pi$}

The statistical model predicts equal branching fractions for $\Lambda_c \to
\Sigma^0 2\pi^+ \pi^-$ and $\Lambda_c \to \Sigma^- \pi^0 2\pi^+$, whereas
the first [$(1.10\pm0.30)\%$] is only about half the second [$(2.1\pm0.4)\%$].
(See Table \ref{tab:S3P}.)  The second process can receive a contribution
from the spectator subprocess $c \to \rho^+ s$ with $\rho^+ \to \pi^+\pi^0$;
the first process has no $\pi^0$.  Suppose (to visualize the effect of
hypothetical resonant substructure) we assume the process were dominated by
the final state $\rho^+ \Lambda(1405) \to (\pi^+ \pi^0)+(\pi^+\Sigma^-,
\pi^0 \Sigma^0,\pi^-\Sigma^+)$.  Then the second, third, and fourth branching
fractions in Table \ref{tab:S3P} would all be 1/3, and the implied total
${\cal B}(\Lambda_c \to \Sigma 3 \pi)$ would be $3 \times (2.1 \pm 0.4)\% =
(6.3 \pm 1.2)\%$, not that far from the value of $(7.3 \pm 2.4)\%$ quoted in
Table \ref{tab:brs}.  Although this is not a complete model for the $\Sigma
3\pi$ final state, it illustrates the importance of experimentally determining
resonant substructure in multibody $\Lambda_c$ final states.

\section{SEMIILEPTONIC $\Lambda_c$ DECAYS \label{sec:sld}}

The only semileptonic $\Lambda_c$ decays that have been reported are
those to a $\Lambda \ell^+ \nu_\ell$ final state.  A hint that there may
be other semileptonic final states is provided by a calculation assuming
pointlike $\Lambda_c$ and $\Lambda$, whose rate is predicted to be
\beq
\Gamma(\Lambda_c \to \Lambda e^+ \nu_e) = \frac{G_F^2 M(\Lambda_c)^5}
{192 \pi^3} f([M(\Lambda)/M(\Lambda_c)]^2) = 2.52 \times 10^{-13}~
{\rm GeV}~,
\eeq
where $f(x) \equiv 1 - 8x + 8x^3 - x^4 + 12x^2 \ln(1/x) = 0.1763$
for $M(\Lambda) = 1115.68$ GeV and $M(\Lambda_c) = 2286.46$ GeV.
The total decay rate of $\Lambda_c$ for a lifetime of 200 fs is
$3.29 \times 10^{-12}$ GeV, so the pointlike prediction corresponds to
a branching fraction of 7.6\%, about twice the observed rate.  This
suggests that a form factor is present, which can be interpreted as
indicating the presence of excited final hadronic states.  A similar
conclusion can be drawn from a quark model cartoon of charm semileptonic
decay \cite{Gronau:2010if}.  With $(m_c,m_s,m_{u,d}) = (1710,536,364)$ MeV
\cite{Karliner:2016zzc}, a slightly higher branching fraction is obtained
(see Appendix), again indicating that semileptonic $\Lambda_c$ decays are not
saturated by the $\Lambda \ell^+ \nu_\ell$ final state.

A calculation in lattice QCD \cite{Meinel:2016dqj} finds ${\cal B}(\Lambda_c
\to \Lambda e^+ \nu_e) = (3.80 \pm 0.19 \pm 0.11)\%$ and ${\cal B}(\Lambda_c
\to \Lambda \mu^+ \nu_\mu) = (3.69 \pm 0.19 \pm 0.11)\%$, where the first error
comes from lattice QCD and the second from the uncertainty in the $\Lambda_c$
lifetime.  The agreement with experiment further confirms the need for a form
factor and, indirectly, hints at a role for excited final states in $\Lambda_c$
semileptonic decays.

The detection of $\Lambda(1405)$ or $\Lambda(1520)$ in the final state of
$\Lambda_c$ semileptonic decays may not be straightforward.  The former
decays only to $\Sigma \pi$, while the latter decays both to $\Sigma \pi$
and to $N \overline{K}$.  Thus branching fractions are spread over many final
states.  There is also a measurement \cite{Vella:1982ei} ${\cal B}(\Lambda_c
\to e^+ + {\rm anything}) = (4.5 \pm 1.7)\%$ from 1982 which needs to be
re-examined if our proposal is to account for a significant portion of the
missing $\Lambda_c$ decays.

\section{CONCLUSIONS \label{sec:con}}

We have investigated $\Lambda_c$ decays from a global standpoint, finding
impressive progress in mapping out branching fractions.  Nevertheless, there
remains the possibility of a shortfall of about 10\%.  We have suggested
that this could be filled by semileptonic decays to excited final states,
not just the $\Lambda$.  To reduce the uncertainty in the total observed
branching fraction, we urge more studies of modes contining neutrons,
greater investigation of the singly-Cabibbo-suppressed modes, and inclusive
studies of $\eta$ and $\eta'$.  Determination of resonant substructure is a
crucial ingredient in filling gaps only partially addressed by an imperfect
isospin statistical model.  The fact that such progress has already been
made for charmed meson decays \cite{PDG} should serve as an encouragement for
similar advances in our understanding of charmed baryon decays.

\section*{ACKNOWLEDGMENTS} 

We thank Roy Briere, Alexander Gilman, Hai-Bo Li, Xiao-Ryu Lyu, Stefan Meinel,
Hajime Muramatsu, Ruth Van de Water, and Charles Wohl for useful
communications.  J.L.R. wishes to thank the Technion for generous hospitality
during the inception of this work.

\section*{APPENDIX}

We give some details of how branching fractions are quoted in Table
\ref{tab:brs} if they are not taken directly from Ref.\ \cite{PDG}.

Modes with $\ok$ are inferred from those quoted in \cite{PDG} for $K_{\rm S}^0$
by multiplying by 2.  We renormalize ${\cal B}(\Lambda_c \to \Sigma^+\pi^0
\pi^0)$, first measured by Belle \cite{Berger:2018pli} using a PDG (2016)
value ${\cal B}(\Lambda_c \to p K^- \pi^+) = 6.35 \pm 0.33 \%$, by the slightly
smaller value of this normalizing branching ratio given in Table
\ref{tab:pkpi}. Using the same normalization for decays to $\Sigma^+\pi^+\pi^-$
and $\Sigma^0 \pi^+\pi^0$ we also include early measurements of these branching ratios.

For the $N \overline{K} \pi$ and
$\Sigma \pi \pi$ modes, all charge states have been measured, so the
experimental totals in Tables \ref{tab:NKP} and \ref{tab:S2P} are
transcribed in Table \ref{tab:brs}.  For the $N \overline{K} 2\pi$ and
$\Sigma 3 \pi$ modes, not all charge states are measured, so totals implied
by the statistical model are averaged and quoted (with a scale factor for
the $N \overline{K} 2\pi$ average) in Table \ref{tab:brs}.
Finally, the modes described in Sec.\ \ref{sec:other} have only one charge
state, which is used to estimate the missing modes, with the inferred
total quoted in Table \ref{tab:brs}.

An alternative way of estimating the total contribution of
singly-Cabibbo-suppressed decays to $\Lambda_c$ branching fractions is
to use free-quark estimates.  For a crude calculation we may consider
only the subprocess $c \to d u \bar d$, neglecting contributions from
$c \to s u \bar s$ by virtue of phase space suppression.  We take effective
quark masses from Ref.\ \cite{Karliner:2016zzc}: $m_c = 1710$ Mev, $m_s = 536$
MeV, $m_{u,d} = 364$ MeV, implying a phase space enhancement of 1.46 for
$c \to d u \bar d$ relative to $c \to s u \bar d$.  The corredponding ratio of
squared CKM matrix elements is $|V_{cd}/V_{cs}|^2 = (0.2265/0.974)^2 = 0.0541$,
implying a total branching fraction for subprocesses dominated by $c \to d u
\bar d$ of $88.6\% \times 1.46 \times 0.0541 = 7.0\%$.  Thus we could be
missing a few $\Delta S = 0$ modes, not to mention those governed by $c \to
s u \bar s$.

In parallel with the lattice QCD calculations mentioned earlier for $\Lambda_c
\to \Lambda \ell^+ \nu_\ell$ \cite{Meinel:2016dqj}, there appeared recently one
for the Cabibbo-suppressed $\Delta S = 0$ processes $\Lambda_c \to n \ell^+
\nu_\ell$ \cite{Meinel:2017ggx}.  The corresponding branching fractions are
displayed in the right-hand column of Table IX.

\end{document}